\documentstyle[12pt]{article}

\begin{document}
{\sf \begin{center} \noindent {\Large \bf Chaotic flows and cosmic dynamos in anisotropic pseudo-Riemannian four-dimensional spacetime}\\[3mm]

by \\[0.3cm]

{\sl L.C. Garcia de Andrade}\\

\vspace{0.5cm} Departamento de F\'{\i}sica
Te\'orica -- IF -- Universidade do Estado do Rio de Janeiro-UERJ\\[-3mm]
Rua S\~ao Francisco Xavier, 524\\[-3mm]
Cep 20550-003, Maracan\~a, Rio de Janeiro, RJ, Brasil\\[-3mm]
Electronic mail address: garcia@dft.if.uerj.br\\[-3mm]
\vspace{2cm} {\bf Abstract}
\end{center}
\paragraph*{}
It is shown that the existence of the cosmic kinematic fast dynamos
in Bianchi type IX rotating cosmological models, faces severe
difficulties, due to the fact that in these models rotation
increases without bounds, which is strictly forbidden by CMB
astronomical date which imposes strong bounds on its rotation with
respect with its rapid expansion. The only way out of this
difficulty is to assume that at least one of the expansion
directions of this anisotropic universe decays as fast as the
amplification rate of primordial magnetic fields. A solution is
found where only one direction of the anisotropic universe expands
while the other two remain constants. We compute an amplification of
the seed magnetic field in the case where Bianchi IX degenerates
into de Sitter metric, fields amplify from $10^{-6}G$ to $10^{-5}G$
in spiral galaxies for a cosmological constant of the order
$|{\Lambda}|<10^{33}s^{-2}$ and considering that the age of universe
of the order of $10^{10}yrs$. Another example is given by the ABC
chaotic flows in the pseudo-Riemannian spacetime representing the
Kasner anisotropic nonsingular universe. \vspace{0.5cm} \noindent
{\bf PACS numbers:} \hfill\parbox[t]{13.5cm}{02.40.Hw-Riemannian
geometries}

\newpage
\section{Introduction}
 Earlier MacCallum \cite{1} has given an outline review of the importance of anisotropic cosmological models of universe, and
 more recently J.D.Barrow \cite{2} has investigate the anisotropic stresses in Einstein general relativistic homogeneous
 models where he showed that Kantowski-Sachs model is unstable
 against perturbations. In most astrophysical dynamos \cite{3} rotation or
 most clearly differential rotation \cite{4} is of utmost importance
 in the amplification process of the magnetic fields in galaxies
 \cite{5} or accretion discs of Kerr black holes \cite{6}. However
 recently , Brandenburg \cite{7} have shown that Cowling
 antidynamo theorem also applies in accretion discs surrounding
 these very compact stellar objects, and dynamo effects are very
 limited. With these physical motivations in this paper we show that
 accretion discs around black holes are not the only astrophysical
 objects where dynamos present problems , but also on the
 cosmological setting cosmic dynamos do not fit well in some anisotropic
 universes as the Bianchi type IX rotating models considered here.
 Actually we show that the existence of dynamos in this universe implies the existence of an unbounded rotation,
  which is forbidden by cosmological data \cite{8}. The only way out
  of this situation, saving the cosmic dynamo existence is to
  consider the isotropization of the model considering the
  irrotational Bianchi IX model or the de Sitter expanding universe
  model. Though great amplification of cosmic primordial magnetic
  field of the order of the several of orders of magnitude appear in
  galactic dynamos, slow dynamos can amplify a magnetic field of the
  order of $10^{-6}G$ to $10^{-5}G$ which happens in spiral
  galaxies. Actually is exactly this amplification that we show to
  be undertaken by the de Sitter cosmic dynamo here. Another way out is to consider turbulent dynamos \cite{9} in
  anisotropic cosmological models. The paper is organised as follows:
  In section II we review the Bianchi IX model geometry in terms of tetrads and differential forms mathematical tools
  the problem, and show that fast decay of the anisotropic dynamo flow can solve.
  In section III we solve the self-induction equation and compute the
  amplification of galactic magnetic fields. In section III we show
  that Arnold-Beltrami-Childress chaotic flows or ABC flows for
  short are compatible with Kasner cosmological background.
  Conclusions are presented in section V.
  \section{Bianchi type IX rotational cosmic dynamos}
 In this section we present a brief review of the mathematical background of cosmological solution of the equations of Einstein theory of
 gravitation. Let us start by  by the Bianchi type IX line element
 \cite{10}
\begin{equation}
ds^{2}=-dt^{2}+g_{ij}{\chi}^{i}{\chi}^{j}\label{1}
\end{equation}
where the indices $(i,j=1,2,3)$ are summed by making use of Einstein
convention of tensor calculus and ${\chi}^{i}$ represent the
synchronous frame system of Cartan differential forms. They are
given explicitly by
\begin{equation}
{\chi}^{1}=-sinx^{3}dx^{1}+sinx^{1}cosx^{3}dx^{2}\label{2}
\end{equation}
\begin{equation}
{\chi}^{2}=cosx^{3}dx^{1}+sinx^{1}sinx^{3}dx^{2}\label{3}
\end{equation}
\begin{equation}
{\chi}^{3}=cosx^{1}dx^{2}+dx^{3}\label{4}
\end{equation}
which after a simple algebra yields the relation
\begin{equation}
\sum({{\chi}^{i}})^{2}=\sum({dx^{i}})^{2}\label{5}
\end{equation}
This last relation is fundamental for the construction of the de
Sitter metric bellow. In terms of the orthonormal tetrad
${\lambda}^{i}={b^{i}}_{s}{\chi}^{s}$ the Bianchi IX rotating metric
becomes
\begin{equation}
ds^{2}=-dt^{2}+{\delta}_{ij}{\lambda}^{i}{\lambda}^{j}\label{6}
\end{equation}
where ${\delta}_{ij}$ is the Kr\"{o}necker delta function. These
models expands or contracts according to the non-vanishing scalar
${\Theta}:= {u^{\alpha}}_{;{\alpha}}$, where $(\alpha=0,1,2,3)$ and
$u^{\alpha}$ is the four-velocity of the cosmic flow. In the
synchronous system, with the help of geodesic equation one can
express the scalar ${\Theta}$ and the rotation tensor
${\omega}_{ij}$ becomes
\begin{equation}
{\Theta}=\frac{{l}_{ij}{u}^{i}{u}^{j}}{u_{0}}-u_{0}l_{kk}\label{7}
\end{equation}
\begin{equation}
{\omega}_{ij}=-\frac{1}{2}{u}^{k}{d^{k}}_{ij}\label{8}
\end{equation}
The metric tensor components $g_{ij}$ are given in terms of triad
coefficients $b_{ij}$ by
\begin{equation}
g_{ij}={b_{i}}^{l}{b_{lj}} \label{9}
\end{equation}
where $b_{ij}$ satisfies the relation $b_{[ij]}=0$. Coefficient
${d^{k}}_{ij}$ is defined as
\begin{equation}
{d^{i}}_{jk}:={b_{i}}^{l}{\epsilon}_{lmn}{b^{-1}}_{mj}{b^{-1}}_{nk}\label{10}
\end{equation}
The coefficients ${l}_{jk}$ are expressed as
\begin{equation}
{l}_{ij}:={\dot{b}}_{(i|l}{b^{-1}}_{l|j)}\label{11}
\end{equation}
the triad $b_{ij}$ is defined as $b_{ij}=b_{ij}(t)$ are solely a
function of time. Here the Levi-Civita symbol ${\epsilon}_{ijk}$,
where  ${\epsilon}_{123}:=1$, is the totally skew symmetric object.
With this mathematical tool we are able to express the vorticity
vector as ${\omega}^{i}={\epsilon}^{ijk}{\omega}_{jk}$. Let us now
consider that the triad $b_{ij}={b^{0}}_{ij}e^{pt}$. The main reason
for this choice is that the magnetic field in tetrad components may
be written as $B^{(i)}={b^{0}}^{is}e^{pt}B_{s}$, which is exactly
the form generally used to investigate fast dynamos \cite{9}.
Substitution of our triad choice into the rotation tensor expression
(\ref{8}) one obtains
\begin{equation}
{\omega}_{ij}=-\frac{1}{2}{u}^{k}{{d^{0}}^{k}}_{ij}e^{pt}\label{12}
\end{equation}
Note from the expression of the magnetic field that the existence of
a kinematic fast cosmic dynamo would demand that $p>0$ and so
expression (\ref{12}) shall tell us that the rotation of the
cosmological fluid rotation increases without bounds as
$t\rightarrow{\infty}$, which unfortunately is strictly forbidden
from avaluable COBE data for expanding universe, which tells us that
rotation of the universe is several orders of magnitude lower than
the expansion of the universe. The only way out of this difficulty
would be to consider that the dynamo flow decays as
$u^{k}=(u^{0})^{k}e^{-pt}$ which would neutralize the fast growing
term $e^{pt}$ in expression (\ref{12}), and we end up with a
constant vorticity for the Bianchi IX model. Let us now consider
that only $u^{z}$ component of the dynamo flow does not vanish and
let us compute the self-induction equation
\begin{equation}
{\partial}_{t}\vec{B}={\eta}{\nabla}^{2}\vec{B}+(\vec{u}.{\nabla})\vec{B}-(\vec{B}.{\nabla})\vec{u}
\label{13}
\end{equation}
By considering a highly conductive cosmic fluid diffusion ${\eta}$
vanishes and this equation reduces to
\begin{equation}
{\partial}_{t}\vec{B}=(\vec{u}.{\nabla})\vec{B}-(\vec{B}.{\nabla})\vec{u}
\label{14}
\end{equation}
to simplify matters we then imagine that the dynamo flow is
homogeneous while the magnetic field also depends on space. This
simplifies equation (\ref{13}) even further to
\begin{equation}
{\partial}_{t}\vec{B}=(\vec{u}.{\nabla})\vec{B} \label{15}
\end{equation}

\section{De Sitter cosmic dynamos in spiral galaxies}
In this section we show that the difficulty with the Bianchi IX
model for the existence of cosmic dynamos in this cosmological
setting does not appear in de Sitter inflationary cosmology
\cite{10}.
 By considering that the four-velocity of the cosmological fluid is
 orthogonal to the $t=constant$, synchronous, space-like
 hypersurface, where $u^{0}=1$ and $u^{i}=0$, so in this frame since
 we are dealing with diagonal metrics, the expression (\ref{8}) tells
 us that the vorticity ${\omega}^{i}=0$. The Friedmann metric
 belongs  to this class of metrics, allows us to consider the de
 Sitter metric
\begin{equation}
ds^{2}=-dt^{2}+e^{pt}[dx^{2}+dy^{2}+dz^{2}]\label{16}
\end{equation}
according to the idea of the last section which allows us to write
the triad ${\lambda}^{(i)}$ as
 \begin{equation}
 {\lambda}^{(i)}={{(b^{0})}^{i}}_{s}e^{pt}{\chi}^{s}\label{17}
\end{equation}
Let us now show that if the homogeneous magnetic field obeys the
magnetic self-induction equation, thus the triad obtained are
exactly  the triad generating de Sitter equation. This is easily
accomplished by considering that homogeneous magnetic fields obey
the self-induction equation
\begin{equation}{\partial}_{t}[{(b)^{i}}_{s}B_{0}]=0
\label{18}
\end{equation}
This equation yields
\begin{equation}({\partial}_{t}{(b)^{is}}) B_{s}+(b)^{is}{\partial}_{t}B_{s}=0
\label{19}
\end{equation}
which is equivalent to
\begin{equation}({\partial}_{t}{(b)^{is}})+(b)^{is}p=0
\label{20}
\end{equation}
where we have used the definition of the magnetic field given in the
previous section. Solving equation (\ref{20}) we obtain the triad
${b^{i}}_{s}={{\delta}^{i}}_{j}e^{pt}$ which yields exactly the de
Sitter triad. Going back to de Sitter cosmology we see that
$p=\sqrt{\frac{\Lambda}{3}}$ where ${\Lambda}$ is the cosmological
constant. Taking the upper limit of $|{\Lambda}|<10^{-35}s^{-2}$ and
the age of universe as $10^{10}yrs$ we are able to think that dynamo
amplified field of de Sitter cosmic dynamo can be approximated as
\begin{equation}
{B_{0}}e^{pt}=B^{0}[1-pt]\label{21}
\end{equation}
where $B_{0}$ is the seed field to be amplified. From the above
data, and assuming that we have a typical seed field of a spiral
galaxy field of $10^{-6}G$ one obtains $10^{-5}G$.
\section{ABC flows in Kasner universe}
In 1981 Arnold, Zeldovich, Ruzmaikin and Sokoloff \cite{11} a
magnetic field in a stationary flow with stretching in a Riemannian
three dimensional space, stationary flow which is exactly the so
called Beltrami 1882 flow also lately investigated by Childress
\cite{9}. In this section, in a certain sense, we extend the ABC
flows defined by the equations
\begin{equation}
u_{x}=A sinz+ C cosy\label{22}
\end{equation}
\begin{equation}
u_{y}=B sinx+ A cosz\label{23}
\end{equation}
\begin{equation}
u_{z}= C siny+ B cosx\label{24}
\end{equation}
to a pseudo-Riemannian anisotropic spacetime called Kasner
cosmological model. Note that \cite{13} the case $A=B=1$ and $C=0$,
represents a fast dynamo. As we shall see bellow our case is $A=C=0$
and $B=1$. To this end we substitute the equations for the flow
$\vec{u}=(u_{x},u_{y},u_{z})$ into the self-induction equation,
assuming now that $\vec{B}=\vec{B_{0}}t^{m}$ where
$\vec{B_{0}}:=B^{y}\vec{j}+\vec{B^{z}}\vec{k}$. This structure is
encoded into the Kasner nonsingular cosmological spacetime
\begin{equation}
ds^{2}= -dt^{2}+dx^{2}+dy^{2}+t^{2}dz^{2}\label{25}
\end{equation}
The equation ${\nabla}.\vec{B}=0$ and self-induction equation
becomes
\begin{equation}
{\partial}_{y}B_{y}+{\partial}_{z}B_{z}=0\label{26}
\end{equation}
\begin{equation}
m{B^{z}}_{0}=-cosx{\partial}_{z}{B^{z}}_{0}\label{27}
\end{equation}
Solution of the last equation yields
\begin{equation}
{B^{z}}_{0}=B_{2}e^{\frac{mz}{cosx}} \label{28}
\end{equation}
substitution of this expression into (\ref{26}) yields
\begin{equation}
{\partial}_{y}B_{y}=-{\partial}_{z}B_{z}=\frac{m}{cosx}{B^{z}}_{0}\label{29}
\end{equation}
which can be easily solved to yield
\begin{equation}
{B^{y}}_{0}=-B_{1}e^{\frac{mz}{cosx}} \label{30}
\end{equation}
where $B_{1}$ and $B_{2}$ are integration constants. This shows that
the magnetic field is amplified if $m>0$ and the flow and the
magnetic field are spatially periodic.
\section{Conclusions}
 In conclusion, we have investigated a chaotic flows, such as ABC chaotic flow, and the cosmic dynamo in Einstein's gravitational equations in four-dimensional
 anisotropic spacetime or in anisotropic pseudo-Riemannian space , which seems to certain extent be a generalization of
 the chaotic non-relativistic ABC flow in a three-dimensional Riemannian space investigated by Arnold and his group.
 Of course other types of chaotic flows in isotropic and anisotropic
 rotating models such as the G\"{o}del model can be investigated
 elsewhere.
 \section*{Acknowledgements}
 Thanks are due to CNPq and UERJ for financial supports.

\end{document}